# Crystallites in Color Glass Beads of the 19th Century and Their Influence on Fatal Deterioration of Glass

E. A. Morozova,[1, 2, a)] I. F. Kadikova,[1, b)] T. V. Yuryeva,[1, c)] V. A. Yuryev [3, d)]

[1]*The State Research Institute for Restoration, Ministry of Culture of Russian Federation, Bldg 1, 44 Gastello Street, Moscow 107114, Russia*
[2]*N. S. Kurnakov Institute of General and Inorganic Chemistry of the Russian Academy of Sciences, 31 Leninsky Avenue, Moscow, 119071, Russia*
[3]*A. M. Prokhorov General Physics Institute of the Russian Academy of Sciences, 38 Vavilov Street, Moscow 119991, Russia*

[a)]Corresponding author: thermochemistme@igic.ras.ru
[b)]kadikovaif@gosniir.ru
[c)]yuryevatv@gosniir.ru
[d)]vyuryev@kapella.gpi.ru

**Abstract.** Glass corrosion is a crucial problem in keeping and conservation of beadworks in museums. All kinds of glass beads undergo deterioration but blue-green lead-potassium glass beads of the 19th century are subjected to the destruction to the greatest extent. Blue-green lead-potassium glass beads of the 19th century obtained from exhibits kept in Russian museums were studied with the purpose to determine the causes of the observed phenomenon. For the comparison, yellow lead beads of the 19th century were also explored. Both kinds of beads contain Sb but yellow ones are stable. Using scanning electron microscopy, energy dispersive X-ray microspectrometry, electron backscatter diffraction, transmission electron microscopy and X-ray powder analysis, we have registered the presence of crystallites of orthorhombic $KSbOSiO_4$ and cubic $Pb_2Sb_{1.5}Fe_{0.5}O_{6.5}$ in glass matrix of blue-green and yellow beads, respectively. Both compounds form at rather high temperatures obviously during glass melting and/or melt cooling. We suppose that the crystallites generate internal tensile strain in glass during its cooling which causes formation of multiple microcracks in inner domains of blue-green beads. We suggest that the deterioration degree depends on quantity of the precipitates, their sizes and their temperature coefficients of linear expansion. In blue-green beads, the crystallites are distributed in their sizes from ~200 nm to several tens of μm and tend to gather in large colonies. The sizes of crystallites in yellow beads are several hundreds of nm and their clusters contain few crystallites. This explains the difference in corrosion of these kinds of beads containing crystals of Sb compounds.

## INTRODUCTION

Glass seed bead embroideries were very popular in the 18th and 19th centuries both in Europe and in Russia.[1,2] Beadworks depicting landscapes, portraits, flowers, battle-pieces, genre scenes, icons were widely used for housewares decoration (Fig. 1). Such handiworks are presented in many arts and historic museums of Europe, North America and Russia among collections of arts and crafts.

The problem in keeping and conservation of these items is acute due to glass corrosion.

It was found that the blue-green (or turquoise) glass seed beads of the early 19th century undergo extreme deterioration in comparison with other kinds of beads.[3,4]

Nowadays, corrosion mechanisms of atmospheric weathered glass have been well studied and this process is known in the literature as crizzling. The glass disease is characterized by the arising of a leached layer at the early stages of corrosion and eventually by the formation of crystalline weathering products, mainly sulphates, carbonates and more rarely nitrates, chlorides and some organic compounds. In this case, glass becomes hazy and cloudy and the surface has a slimy or soapy feel. As a rule the degradation process leads to rupture of glass since appearing stress in glass induces its cracking.[5,6]

The detailed examination of beadworks with corroded turquoise glass beads identified the following feature: there are both intact blue-green glass beads and degraded ones on the same beaded embroideries of the 19th century. Moreover, the beads at different stages of corrosion are situated side by side (Fig. 2 (a)). Being used in household beaded items, beads have obviously been subjected to the same impact of external factors but corroded to different degree for the same time. So, it should be concluded that glass corrosion in the case of the blue-green beads has another reason. The understanding of how unstable glass degrades would help develop a long-term conservation technique.

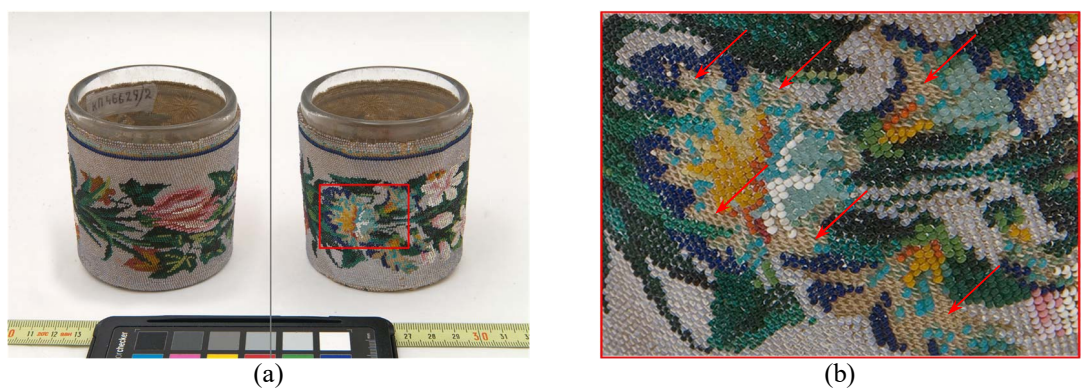

**FIGURE 1.** Beaded glass-holder of the 19th century from the collection of the Museum of A. S. Pushkin, Moscow (a) and its fragment (b) before restoration; numerous gaps of turquoise glass seed beads are seen in panel (b), some of them are pointed with arrows.

## SAMPLES AND METHODS

Samples of cloudy blue-green and yellow glass seed beads were obtained during restoration of secular beaded items of the 19th century from museum collections of Russia (Fig. 1). During the study of the blue-green beads, samples with different deterioration degrees were explored (Fig. 2 (b)–(h)).

Before all experiments, samples were washed with high purity isopropyl alcohol ([$C_3H_7OH$] > 99.8 wt. %) at 40°C for 20 minutes in a chemical glass placed into an ultrasonic bath (40 kHz, 120 W).

Scanning electron microscopy (SEM) studies were performed using Tescan Vega-II XMU and Carl Zeiss NVision 40-38-50 microscopes in the modes of secondary electrons (SE) and backscattered electrons (BSE).

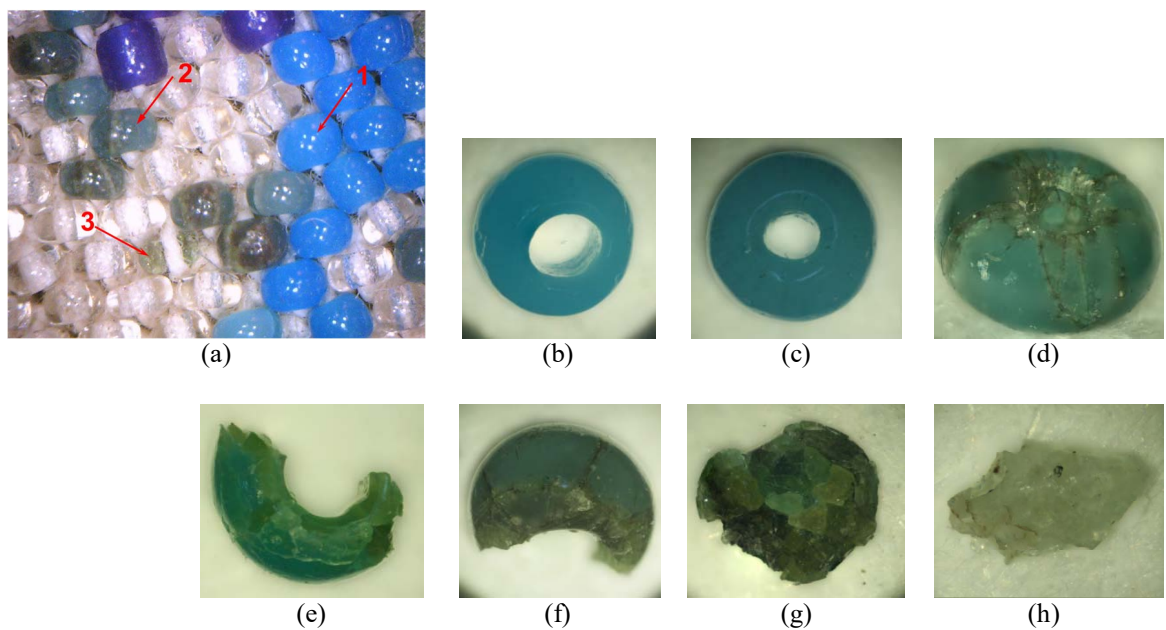

**FIGURE 2.** (a) Macro photograph of a damaged part of a beaded glass-holder; blue-green glass beads at different degrees of degradation (intact, severely cracked with color change and partially broken) are pointed by numbers 1, 2 and 3, respectively. (b)–(h) Micro photographs of blue-green beads at different stages of deterioration in the reflected polarized light: firstly, cracks appear in the blue beads. While cracking increases, the color starts to change into a greenish or yellowish tone. Further degradation is accompanied by bead discoloration and formation of a grainy structure. Such beads are very fragile so they eventually break into fragments.

Inca Energy 450 energy dispersive X-ray spectrometers with Inca DryCool detectors were used for microanalysis of elemental composition.

The X-ray phase analysis was performed by means of the Debye-Scherrer powder diffraction method using the Bruker D8 Advance diffractometer at non-monochromatic $CuK_\alpha$ band ($CuK_{\alpha1,2}$, $\lambda = 1.542$ Å); the diffraction patterns were scanned in a 2θ interval from 5 to 75º with the steps of 0.02º; the data acquisition time per one point was 9.0 sec. PDF-2 Powder Diffraction Database was used for the phase composition analysis. In addition, Crystallography Open Database (COD) was also used for the data analysis.

Additionally, for the direct phase analysis of inclusions in glass matrix, a Nordlys II S electron backscatter diffraction (EBSD) detector and Channel 5 software (Oxford Instruments HKL) were used.

Elemental composition of all samples was also analyzed using a M4 TORNADO micro-X-ray fluorescence (XRF) spectrometer (Bruker).

Composition of inclusions in glass matrix was explored by means of transmission electron microscopy (TEM) including scanning TEM (STEM). A Carl Zeiss Libra-200 FE transmission electron microscope was used. Samples for TEM were prepared using Model 1010 ion mill (E. A. Fischione Instruments) at ion accelerating voltage of 4 kV.

## RESULTS AND DISCUSSION

### Blue-Green Glass Beads

*SEM and EDX*

SEM images and EDX spectra are represented in Fig. 3. Numerous faceted particles were found at different points of the studied samples during the detailed analysis of glass surface around the cracks and far from them.

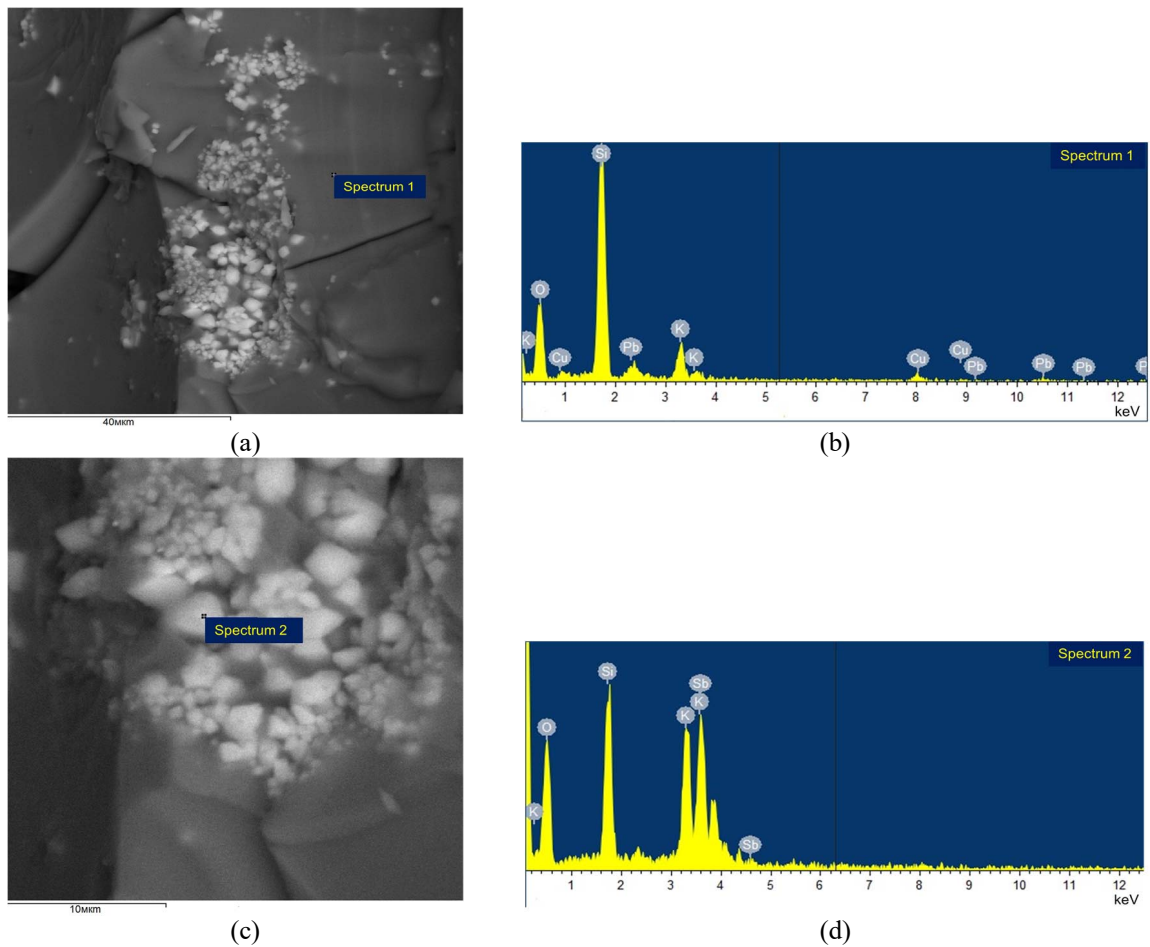

**FIGURE 3.** (a), (c) SEM BSE images of glass surface area with colonies of some crystallites (after fragmentation of turquoise bead at the final degradation stage); (b), (d) EDX spectra obtained at points indicated in panels (a) and (c) (Spectrum 1 and Spectrum 2).

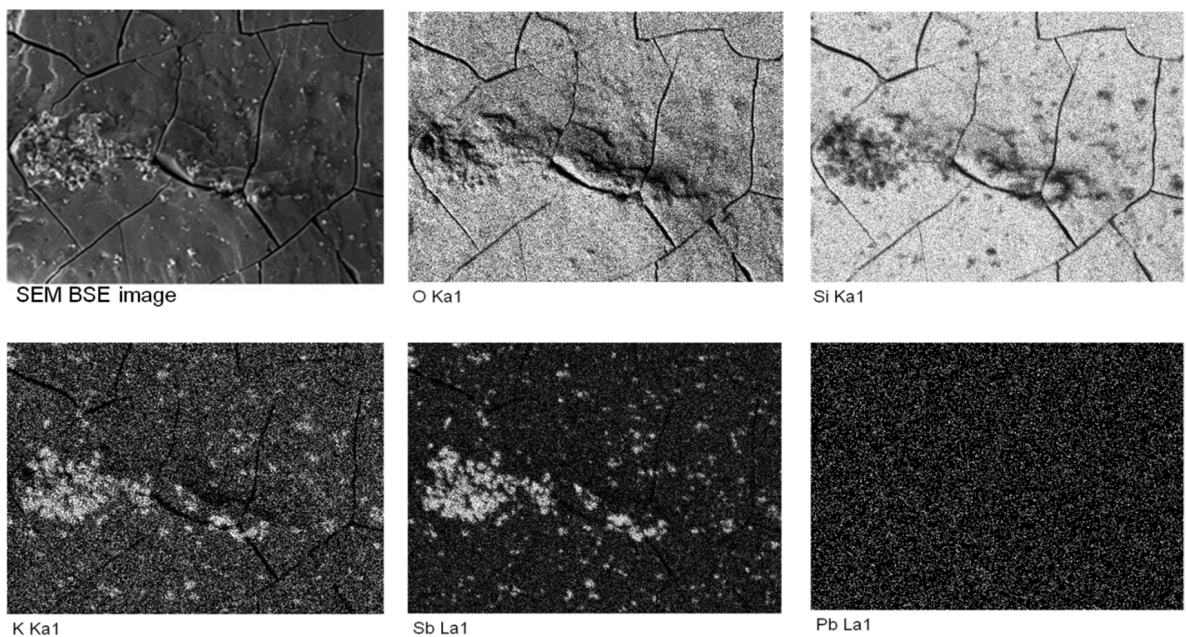

**FIGURE 4.** Maps of some chemical elements obtained using EDX during the investigation of a selected scanned area.

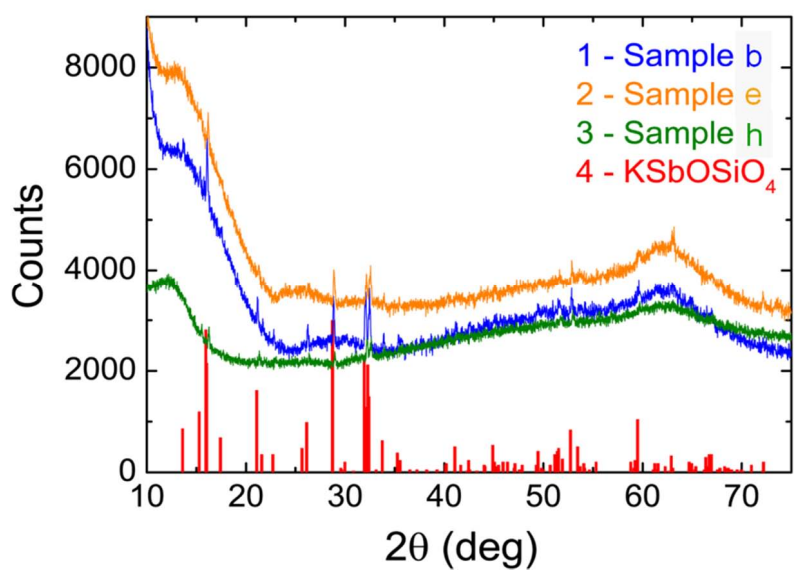

**FIGURE 5.** X-ray powder patterns of bead samples at different degradation stages (see panels (b), (e) and (h) in Fig. 2) demonstrating the presence of the orthorhombic $KSbOSiO_4$ phase.

The investigation of the composition of glass bulk by the use of EDX has allowed us to establish that turquoise seed beads are made of lead-potassium glass (Fig. 3, spectrum 1). Such elements as copper and aluminum were also registered.

The crystalline inclusions found in blue-green beads at different phases of corrosion—intact, slightly cracked, severely cracked and changed the color or discolored, partially or completely fragmented—contain potassium, antimony, silicon and oxygen in their composition (Fig. 3, spectrum 2).

To visualize chemical elements distribution using EDX, we have plotted their spatial maps (Fig. 4).

The following features have been observed: (i) regions of decreased fraction of silicon coincide with the Sb-rich precipitates; (ii) regions of increased fractions of antimony and potassium also coincide with the Sb-rich precipitates; (iii) no deviations of oxygen fraction from its mean value are observed; (iv) other elements registered in EDX spectra do not demonstrate any reliable deviations from their mean values in glass.

*X-ray Powder Diffraction*

The X-ray powder diffraction analysis of blue-green glass beads at different stages of deterioration was carried out in order to define the phase composition of the observed crystallites. Orthorhombic potassium-antimony silicate KSbOSiO$_4$ (further KSS) was identified in all samples [symmetry $Pna2_1$ (33), PDF-2 Card # 01-080-1595] (Fig. 5).

Some reflexes detected in diffraction pattern have not been identified. We suppose that they relate to a crystalline phase of SiO$_2$; it requires further clarification, however. Nevertheless, it should be noted that the devitrification of glass under study is not observed even over the time far exceeding human lifespan. It correlates with the properties of lead oxide based glasses that are applied in optical and electronic devices and characterized by high refractivity, chemical durability, low melting point and stability against glass crystallization.[7]

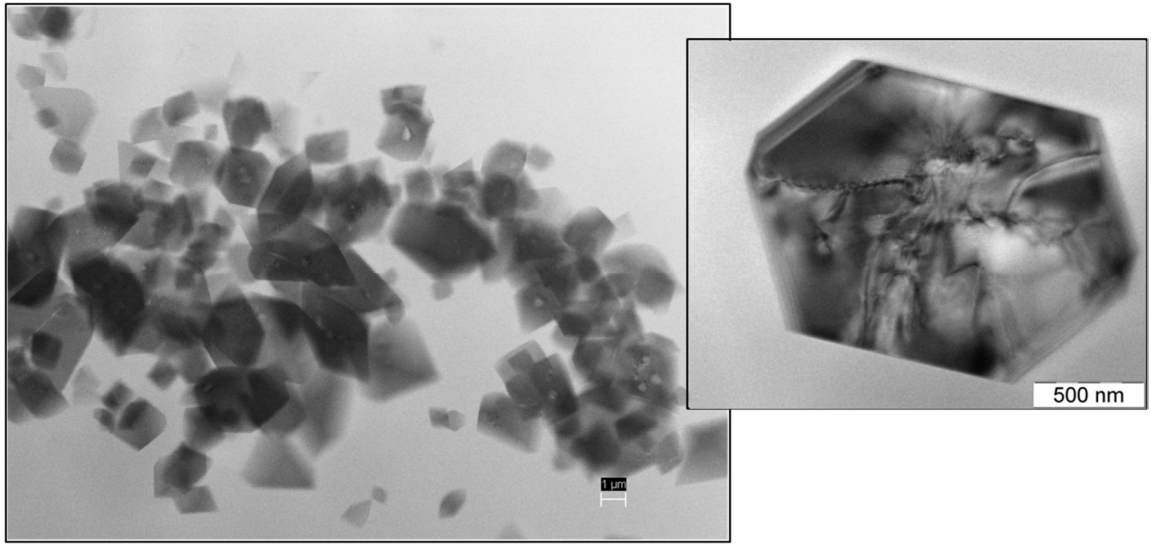

**FIGURE 6.** STEM image (left) of a KSS cluster and TEM image (right) of a typical KSS crystallite.

*TEM and STEM*

Using transmission electron microscopy, we managed to estimate the sizes of the KSS crystallites, which vary from nearly 200 nm to several tens of μm. This compound forms whole colonies (Fig. 6).

EDX analyses of the crystallites have shown that they are composed of oxygen, silicon, potassium and antimony.

*EBSD (Transmission)*

Electron backscatter diffraction enables one to explore separate crystallites in contrast to X-ray powder diffraction. Using of EBSD, we managed to confirm that namely orthorhombic KSS compose the observed crystallites in turquoise lead-potassium glass (Fig. 7).

We suppose that these crystallites play a key role in fatal deterioration of turquoise glass beads.

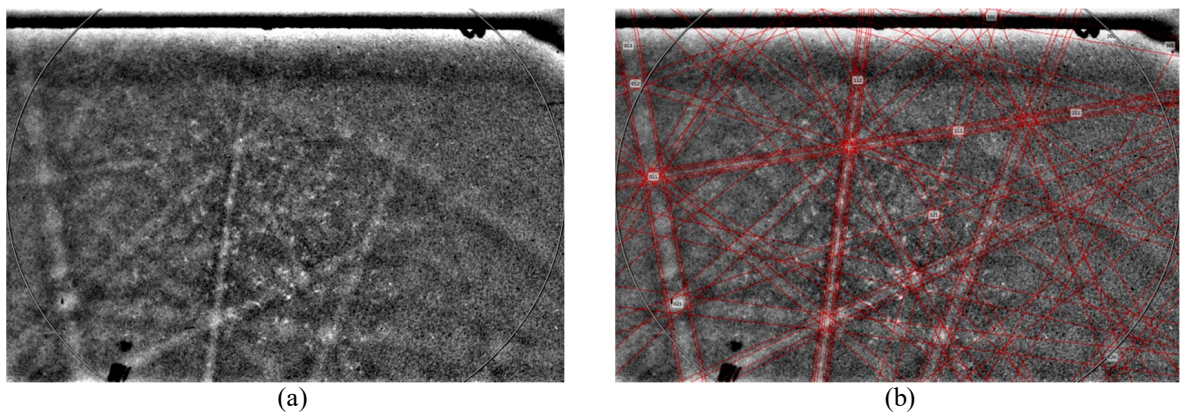

(a)          (b)

**FIGURE 7.** Kikuchi patterns observed in blue-green glass beads (a) and assigned to orthorhombic KSbOSiO$_4$ with the analysis system (b).

The presence of antimony in glass may be connected either with its entering as impurity from main components[8] or with its intentional addition as an opacifier.[9] The crystals formation in glass matrix was associated with the glass manufacture process. Orthorhombic KSS is known to form at temperatures higher than 1100ºC. At lower temperatures, the tetragonal KSS is formed.[10] The latter KSS phase has not been found in our samples, therefore, we believe that the detected KSS crystallites appeared at temperatures higher than 1100ºC during glass melting and bubbling.

The effect of these crystals on glass corrosion process may be described in the following way. Once formed, the KSS crystallites generate internal tensile strain in the cooled glass because of some difference in their linear thermal expansion coefficients. It causes the formation of multiple microcracks in inner domains of beads. SEM images presented at Fig. 8 illustrate the above-mentioned statement: cracks, which look in the micro photograph as if they cover the surface of a bead hole, in reality do not reach the glass surface and are located in the bead internal domain.

The formation of multiple microcracks eventually gives rise to glass rupture and formation of heterogeneous sand grains.

The diffusion of most of the elements induced by tensile strain may explain changes in glass color.

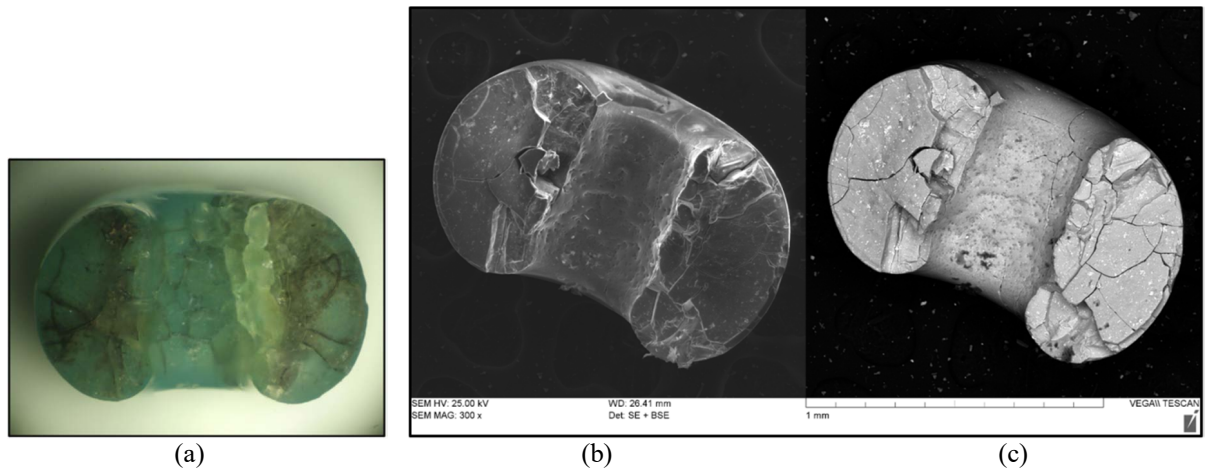

(a) (b) (c)

**FIGURE 8.** (a) Micro photograph of a degraded blue-green glass bead in the reflected polarized light, (b) SEM SE and (c) SEM BSE images of the same sample.

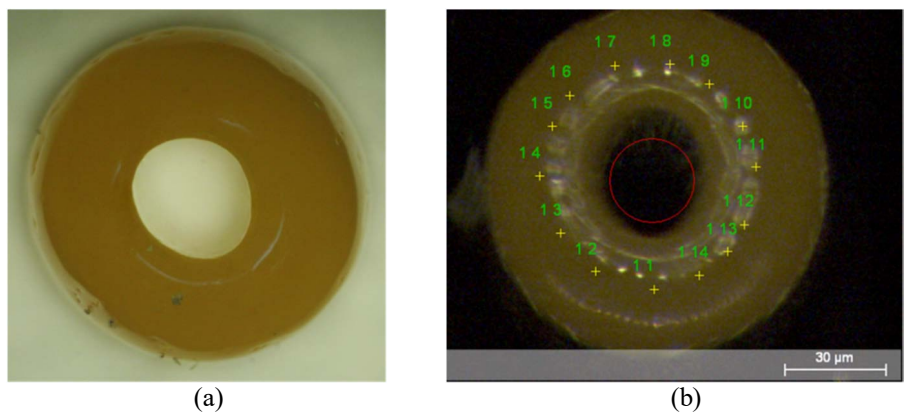

(a) (b)

**FIGURE 9.** Micro photograph of yellow glass bead in the reflected polarized light (a) and an image of the sample with indication of the points at which XRF measurements were carried out.

**TABLE 1.** Elemental composition of yellow bead glass, at. %

|  | Al | Si | P | S | K | Ca | Mn | Fe | Cu | Zn | As | Sb | Pb |
|---|---|---|---|---|---|---|---|---|---|---|---|---|---|
| Mean value | 1.47 | 52.67 | 0.07 | 7.30 | 0.20 | 2.34 | 0.10 | 0.71 | 0.20 | 0.04 | 0.54 | 0.84 | 33.52 |

# Yellow Glass Beads

As distinct from blue-green glass seed beads, the yellow ones distinguish by very high corrosion resistant.

*XRF*

The study by the use of X-ray fluorescence microspectrometry has shown that the yellow beads are made of lead glass (Fig. 9, Table 1). The content of lead in them is significantly higher than in blue-green lead-potassium glass. Like turquoise glass, the yellow one also contains antimony but in much less proportion. According to our previous work,[4] like in the turquoise beads, antimony is contained only in inclusions, not in glass matrix.

*TEM and STEM*

Using transmission electron microscopy, we managed to determine the composition of these inclusions and to estimate their sizes. STEM image presented at Fig. 10 (a) demonstrates that these inclusions are clusters of small crystallites. Their sizes vary from a few hundreds of nanometers to about 1 μm. According to EDX analysis, the crystallites consist of lead, iron, antimony and oxygen (Fig. 10 (b)). The copper detected using EDX is from the sample holder material.

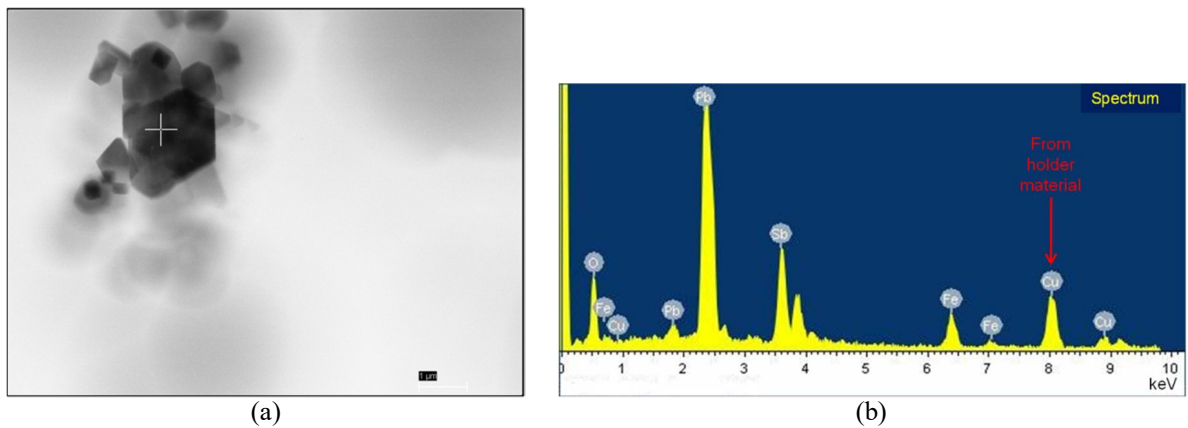

(a) (b)

**FIGURE 10.** (a) STEM image of an inclusion in glass matrix of a yellow bead, EDX analysis point is marked by a cross; (b) EDX spectrum obtained at this point.

*EBSD (Reflection)*

Kikuchi lines shown in Fig. 11 prove the crystallinity of the inclusions and identify them as cubic lead iron antimony oxide $Pb_2Sb_{1.5}Fe_{0.5}O_{6.5}$.

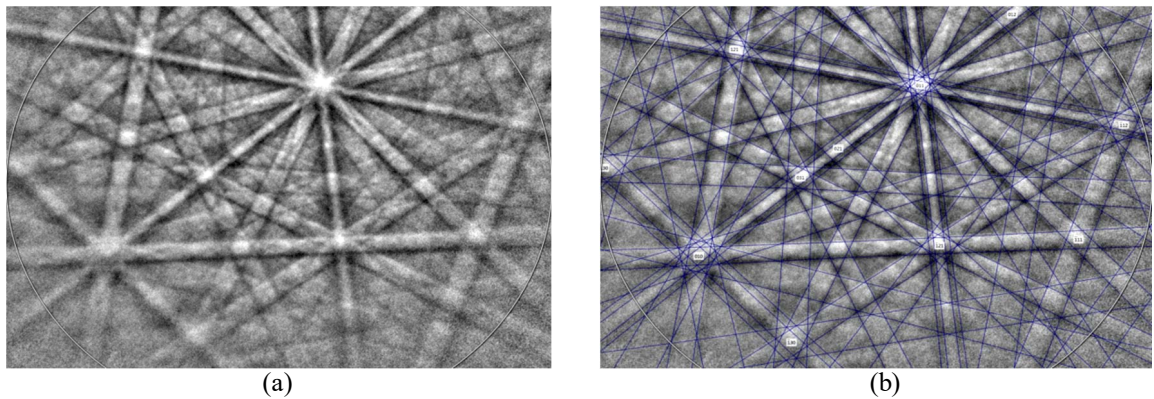

(a) (b)

**FIGURE 11.** Kikuchi patterns observed in yellow glass beads (a) and assigned to cubic $Pb_2Sb_{1.5}Fe_{0.5}O_{6.5}$ with the analysis system (b).

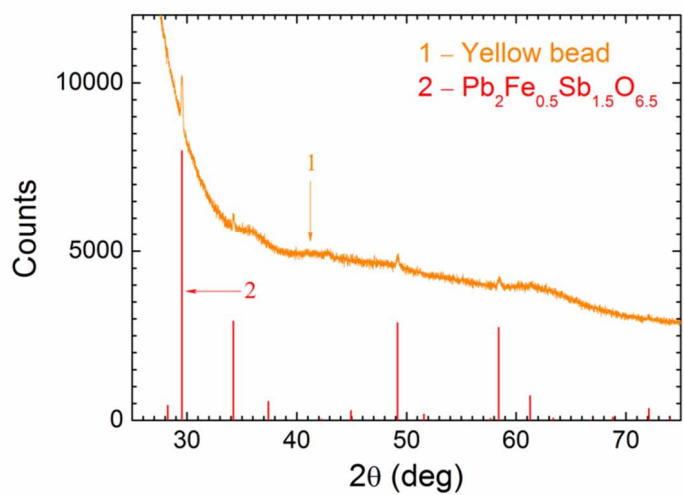

**FIGURE 12.** X-ray powder pattern of a yellow glass bead (Fig. 9) demonstrating the presence of the cubic $Pb_2Sb_{1.5}Fe_{0.5}O_{6.5}$ phase in glass.

*X-ray Powder Diffraction*

The presence of this substance was also determined by X-ray powder diffraction analysis of a yellow bead powder. The diffraction pattern (Fig. 12) corresponds to cubic $Pb_2Sb_{1.5}Fe_{0.5}O_{6.5}$ (symmetry *Fd-3m* (227), PDF-2 Card # 01-077-2454).

This compound is a derivative of Naples yellow ($Sb_2O_5 \cdot nPbO$) with some iron atoms replacing ones of antimony. Naples yellow is a pigment widely used in the 18th and 19th centuries in paintings.[11]

## CONCLUSION

The colonies of micro and nanocrystallites of orthorhombic $KSbOSiO_4$ (KSS) have been detected in glass bulk of blue-green (turquoise) lead-potassium glass beads of the 19th century manufacture.

We believe that the presence of KSS crystallites and their clusters in glass matrix plays a key role in corrosion process proceeding that starts internally rather than from the surface as is common for glass objects.

The formation of KSS crystalline particles is most probably connected with glass melt producing (during glass bubbling and/or melt cooling). Tensile strain arising in the glass matrix during cooling causes glass cracking and eventually its separation into the fragments. The strain-induced diffusion of most of the elements presenting in glass composition may explain changes in glass color of blue-green beads.

We have also detected $Pb_2Sb_{1.5}Fe_{0.5}O_{6.5}$ nanocrystallites in stable yellow lead glass beads. The number density and the sizes of these crystallites are much less than those of the KSS crystallites in turquoise lead-potassium glass, they do not form the colonies; internal microcracks also has not been observed in this glass. This may explain the stability of yellow lead glass to internal corrosion.

The addition of lead oxide to the glass matrix of both kinds of seed beads has increased the resistance of the glass to the environment conditions and prevented the process of glass devitrification.

## ACKNOWLEDGEMENTS

The research was funded by the Russian Science Foundation (grant No. 16-18-10366).

The work was carried out under the Collaboration Agreement of the State Research Institute for Restoration and Prokhorov General Physics Institute of RAS and under the Collaboration Agreement of the State Research Institute for Restoration and Kurnakov Institute of General and Inorganic Chemistry of RAS.